\documentclass[10pt,twocolumn,letterpaper]{article}

\usepackage{iccv}
\usepackage{times}
\usepackage{epsfig}
\usepackage{graphicx}
\usepackage{amsmath}
\usepackage{amssymb}

\usepackage{algorithm}
\usepackage[noend]{algpseudocode}

\usepackage{xcolor}
\usepackage{listings}

\definecolor{mGreen}{rgb}{0,0.6,0}
\definecolor{mGray}{rgb}{0.5,0.5,0.5}
\definecolor{mPurple}{rgb}{0.58,0,0.82}
\definecolor{backgroundColour}{rgb}{0.95,0.95,0.92}

\lstdefinestyle{CStyle}{
    backgroundcolor=\color{backgroundColour},   
    commentstyle=\color{mGreen},
    keywordstyle=\color{magenta},
    numberstyle=\tiny\color{mGray},
    stringstyle=\color{mPurple},
    basicstyle=\footnotesize,
    breakatwhitespace=false,         
    breaklines=true,                 
    captionpos=b,                    
    keepspaces=true,                 
    numbers=left,                    
    numbersep=5pt,                  
    showspaces=false,                
    showstringspaces=false,
    showtabs=false,                  
    tabsize=2,
    language=C
}

\usepackage{mdframed}
\usepackage{lipsum}
\newmdtheoremenv{theo}{Theorem}

\usepackage[breaklinks=true,bookmarks=false]{hyperref}

\iccvfinalcopy 

\def\iccvPaperID{****} 

\ificcvfinal\fi

\begin{document}

\title{Structure-based Computer Without Using Transistors}

\author{Jonghyeon Lee\\
Sejong Academy of Science and Arts\\
Sejong, Korea\\
{\tt\small manatee02@sasa.hs.kr}
\and
Taewon Kang\\
Sejong Academy of Science and Arts\\
Sejong, Korea\\
{\tt\small itschool@itsc.kr} \\
}


\maketitle
\ificcvfinal\thispagestyle{empty}\fi

\begin{abstract}
   The commercialization of transistors capable of both switching and amplification in 1960 resulted in the development of second-generation computers, which resulted in the miniaturization and lightening, while accelerating the reduction and development of production costs. However, the self-resistance and the resistance used in conjunction with semiconductors, which are the basic principles of computers, generate a lot of heat, which results in semiconductor obsolescence, and limits the computation speed (Clock rate). In implementing logic operation, this paper proposes the concept of Structure-based Computer which can implement NOT gate made of semiconductor transistor only by Structure-based twist of cable without resistance. In Structure-based computer, the theory of 'inverse signal pair' of digital signals was introduced so that it could operate in a different way than semiconductor-based transistors. In this paper, we propose a new hardware called Structure-based computer that can solve various problems in semiconductor computers only with the wiring structure of the conductor itself, not with the silicon-based semiconductor. A USB-type Structure-based computer prototype has been built, and a logical cloning method using CPU Clock is proposed to avoid the risk of current being reversed by cloning logical values. Furthermore, we propose a deep-priority exploration-based simulation method that can easily implement and test complex Structure-based computer circuits. Furthermore, this paper suggests a mechanism to implement optical computers currently under development and research based on structures rather than devices.
\end{abstract}

\section{Introduction}
The structural computers produced in this paper are based on Boolean Algebra, a system commonly applied to digital computers. Boolean algebra is a concept created by George Boole (1815-1854) of the United Kingdom that expresses the truth and falsehood of logic 1 and 0, and mathematically describes digital electrical signals. The concept of logical aggregates defined in Boolean algebra has become the basis for hardware devices such as ALU, CLU, RAM, and so on. Structural computers in this paper were also designed to perform logical operations using digital signals of 1 and 0. Logic circuits are the units in which logical operations are performed, and there are AND, OR, and NOT.

Of these, the NOT gate in the computer we use today is based on transistors. The advantage of transistors is that they can differentiate between signal and power and perform switching and amplification at the same time. On the other hand, more heat is generated compared to passing through a conductor of the same length, which causes semiconductors to age and limits the number of clocks.

To solve the various problems of the semiconductor computer mentioned above, this paper tries to present the concept of Reverse-Logic pair of digital signals and double-pair-based logic operation techniques on which structural computer hardware is based.

This introduction introduces the concept of reverse signal pair\cite{phoneworks} (reverse-Logic pair) of digital signals, which is a method for solving the problem of heating, aging, and computation speed of NOT operations.

Expressing 1 as an inverted signal pair, it appears as an ordered pair of two auxiliary signals, each with a signal of one or zero, as shown in (10). Similarly, zeros are expressed in sequence pairs (0,1).

In other words, for any digital signal A, 
\begin{equation}
    A=(\alpha,\beta)\rightarrow\ |A|=\alpha,\ \ \ \ \beta= \sim \alpha\ \nonumber
\end{equation}

it can be expressed as (an expression) and $\alpha$ is defined as a true signal for logical A, $\beta$ as an inverted signal. Using this, the logical operation of two signals A and B is expressed as follows.

\begin{equation}
    for\ \forall A,B\in \{ \left(1,0\right),\left(0,1\right) \} \nonumber 
\end{equation}
\begin{align}
    NOT\ A &= \sim \left(\alpha_A,\beta_A\right) \nonumber \\
    &= \left(\sim \alpha_A, \sim \left(\sim \alpha_A\right)\right) \nonumber \\
    &= \left(\beta_A,\alpha_A\right) \nonumber \\
    A\ AND\ B &= \left(\alpha_A\land\alpha_B,\sim\left(\alpha_A\land\alpha_B\right)\right) \nonumber \\
    &=\left(\alpha_A\land\alpha_B,\sim\alpha_A\vee\sim\alpha_B\right) \nonumber \\
    &=(\alpha_A\land\alpha_B,\beta_A\vee\beta_B) \nonumber \\
    A\ OR\ B &= \left(\alpha_A\vee\alpha_B,\sim\left(\alpha_A\vee\alpha_B\right)\right) \nonumber \\
    &=\left(\alpha_A\vee\alpha_B,\sim\alpha_A\land\sim\alpha_B\right) \nonumber \\
    &=(\alpha_A\vee\alpha_B,\beta_A\land\beta_B) \nonumber
\end{align}

As shown in the above method, logical aggregates can be constructed with structural wiring if digital signals are computed in pairs of inverted signals. Especially for the NOT gate, you can twist the $\alpha$ line and the $\beta$ line once, making it much simpler to operate than a semiconductor-based transistor that uses a traditional semiconductor element and a pore.

In addition, cables were measured in pairs rather than in pairs to enable serial connection when AND operations were performed. The $\alpha$ signal and $\beta$ signal have values of 0 and 1, depending on the connection state of the wire.

\begin{figure}[h]
\begin{center}
\includegraphics[width=4.6cm]{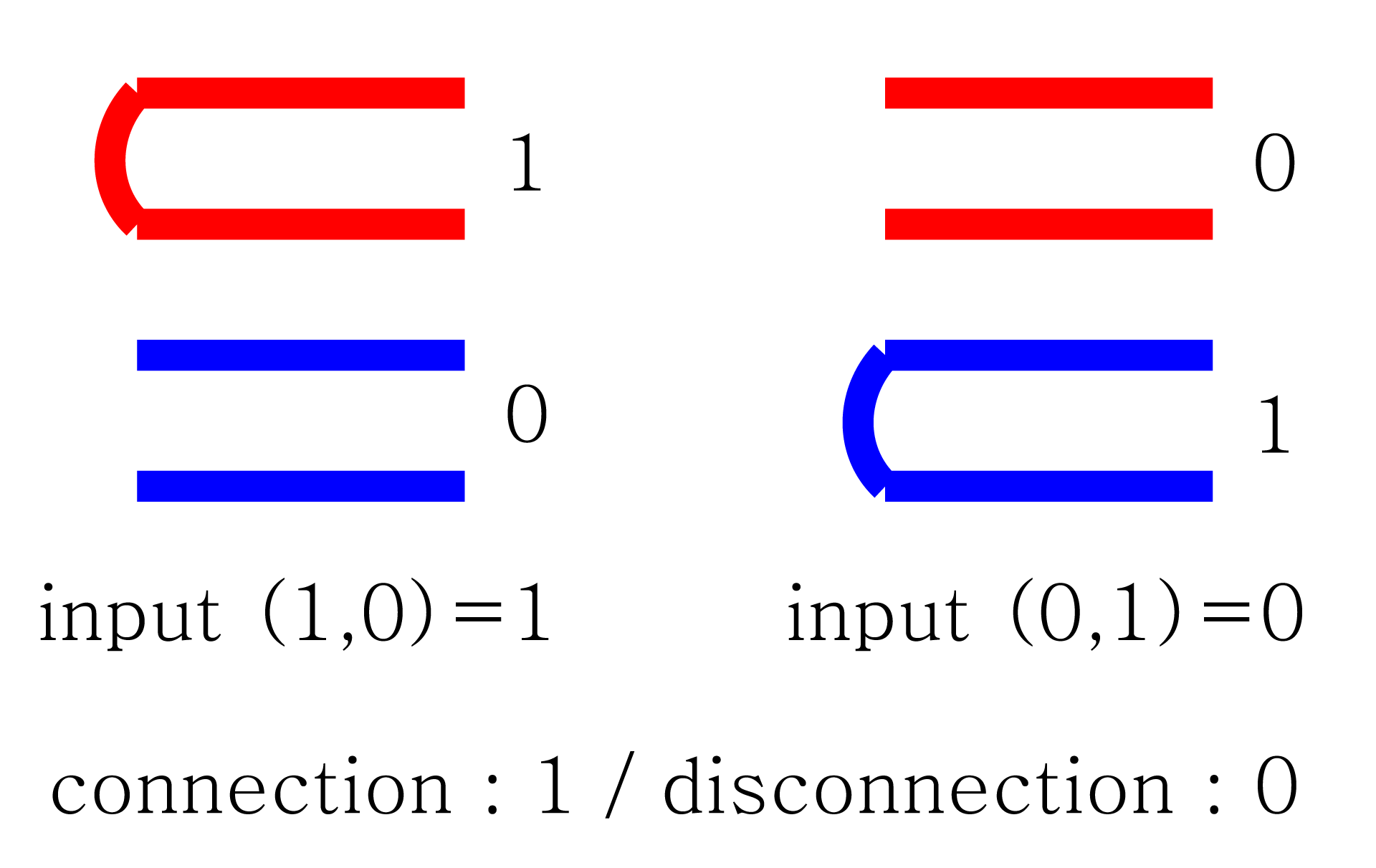}
\end{center}
\caption{\bf Double Pair-based Logical operation input method } \label{Fig01}
\end{figure}

If a pair of lines of the same color is connected, 1, if broken, the sequence pair of states of the red line ($\alpha$) and blue line ($\beta$) determines the transmitted digital signal. Thus, signal cables require one transistor for switching action at the end.

When introducing the concept of an inverted signal pair of digital signals into a structural computer, the signals are paired, so a total of four wires are required to process the two auxiliary signals. This is defined as a double pair-based logical operation and is as follows in \textbf{Fig~\ref{Fig01}}.

\section{Methods}
\subsection{Structure-Based Logic: 4-pin based logic}

As shown above, the two inverted connection states, i.e. the relationship between two inverted signal pairs, are called two pairs-based (double pair-based), and the following 4-pin based logic. As noted in the introduction, the advantage of this system lies in being able to implement logical NOT gates without transistors. Let's look at the structural transformation that intersects in a diagonal shape as shown below.

\begin{figure}[h]
\begin{center}
\includegraphics[scale=0.2,bb=0 0 600 328]{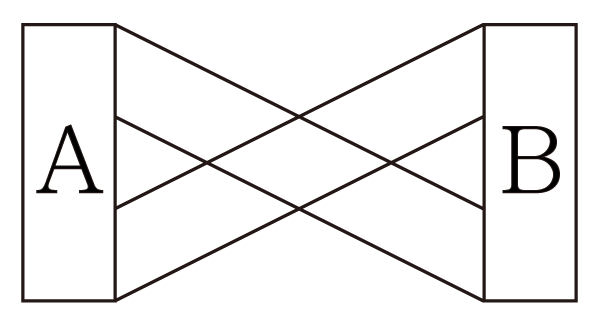}
\end{center}
\caption{\bf 4-pin based NOT gate}\label{Fig02}
\end{figure}

Assuming that the above structure passes through any 4-pin based logic signal, output B is a pair of signals with the true value ($\alpha$) of input A and the inverted ($\beta$) reversed. This means $B= \sim A$, which is demonstrated through a logical formula as follows:

\begin{align}
A=(a,b)&=(a,\sim a) \nonumber \\
\text{because } b &= \sim a \nonumber \\
B=f(A) &= (b,a)=(\sim a,a) \nonumber \\
|B| &= \sim a =~|A| \nonumber
\end{align}

So let's now implement logical (AND) and logical (OR) operations as structural transformations. The figure below shows a link that allows the logical (AND) computation of the signal input to A and B.

\begin{figure}[h]
\begin{center}
\includegraphics[scale=0.15,bb=0 0 600 600]{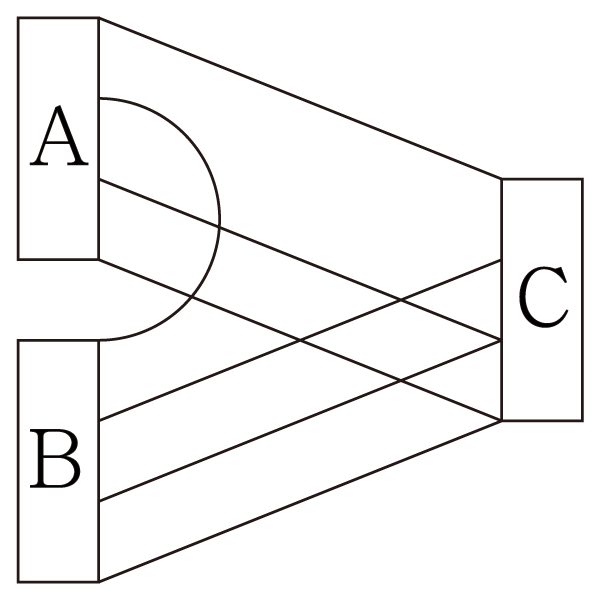}
\end{center}
\caption{\bf 4-pin based AND gate}\label{Fig03}
\end{figure}

As shown in \textbf{Fig.~\ref{Fig03}}, true values of A and B are linked in series and inverted values in parallel. This is a connection using the concept of ‘Keep the reverse signal pair through OR operation with the half value when performing an AND operation between the true values as demonstrated by the logic presented in Induction’ as a means to maintain the reverse signal pair of outputs.

\begin{figure}[h]
\begin{center}
\includegraphics[scale=0.15,bb=0 0 600 600]{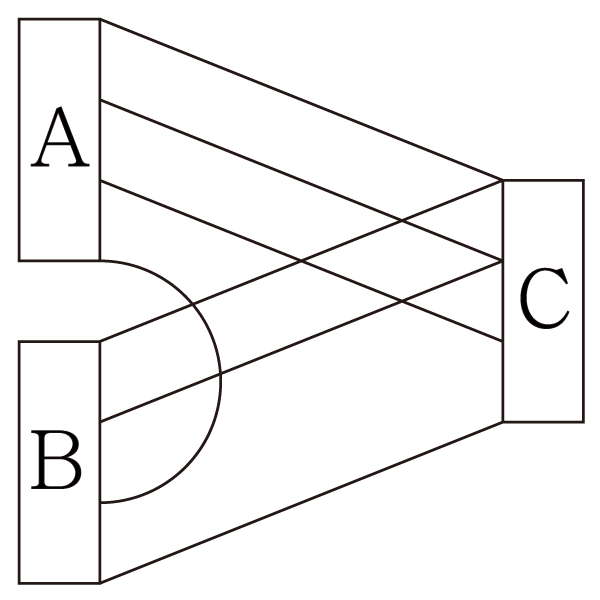}
\end{center}
\caption{\bf 4-pin based OR gate}\label{Fig04}
\end{figure}

\textbf{Fig.~\ref{Fig04}} shows if the OR gate is configured in the same way. In contrast to AND gate, there is a parallel and inverted connection between true values, which is also a connection using the concept of ‘maintaining an inversion signal pair through OR operation when performing an AND operation between true values, as in AND gate. For this reason, the OR gate of the structural computer looks like an AND gate upside down, and I will refer to 2.3 for the mathematical principles of this property.

\begin{figure}[h]
\begin{center}
\includegraphics[scale=0.15,bb=0 0 600 600]{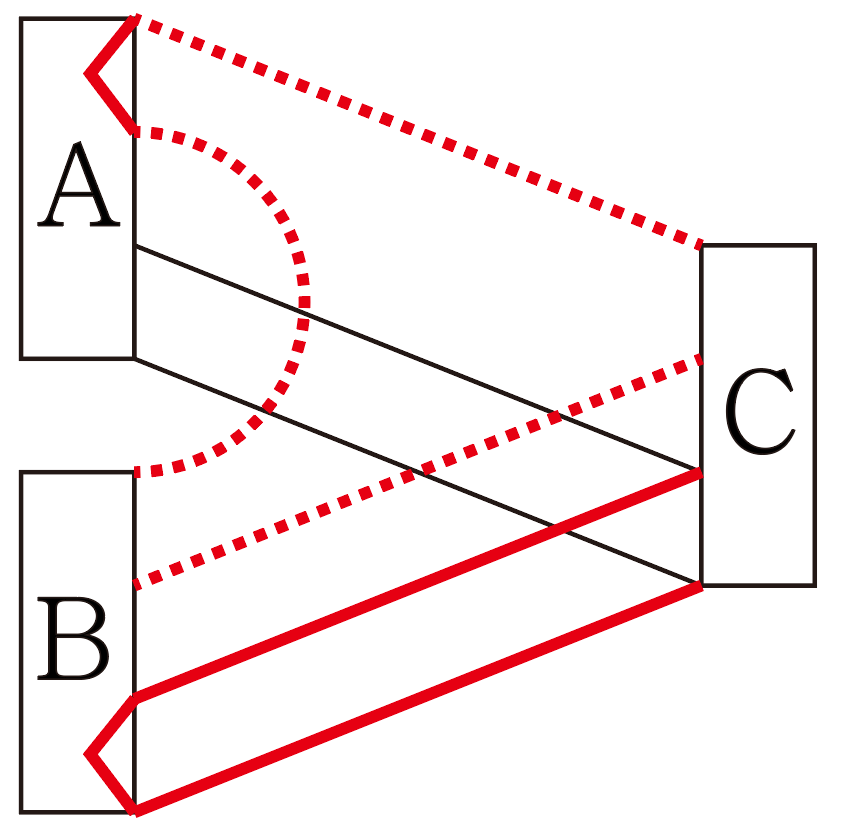}
\end{center}
\caption{\bf 4-pin based AND gate inputs A=1, B=0}\label{Fig05}
\end{figure}

For example, \textbf{Fig.~\ref{Fig05}} shows that the top two of the four pins of C (output) are not connected and the bottom two are connected by inputting A=1, B=0 into the AND gate.

The graph shown in \textbf{Fig.~\ref{Fig05}} is a schematic diagram showing the true value of output C and the state of connection of the inverted value pin when input values A=1, B=0 are entered into the 4-pin AND gate. Since the true value of C is unconsolidated and the inverted value is connected, the logical value of C is zero.

\subsection{Structure-based Logic: 3-pin based logic}
The structural computer used an inverted signal pair to implement the reversal of a signal (NOT operation) as a structural transformation, i.e. a twist, and four pins were used for AND and OR operations as a series and parallel connection were required.

However, one can think about whether the four pin designs are the minimum number of pins required by structural computers. In other words, operating a structural computer with a minimal lead is also a task to be addressed by this study because one of the most important factors in computer hardware design is aggregation.

Let's look at the role of the four pins that transmit signals in a 4 pin based signal system. Four pins are paired into two pairs, each representing/delivering true and inverted values as a connection state.

When checking the output, place a voltage on one of the two wires in a pair and ground the other. In this case, the study inferred that of the four wires, two wires acting as ground can be replaced by one wire, and based on this reasoning, the method in which the 4 pin signal system can be described as \textbf{3-pin based logic} as the same 3 pin signal system.

As mentioned above, a 3-pin based logic consists of a ground cable in the center and two signal lines representing true and inverted values above and below, and is capable of operating NOT, AND and OR operations through the structural transformations shown below.

\begin{figure}[h]
\begin{center}
\includegraphics[scale=0.15,bb=0 0 600 328]{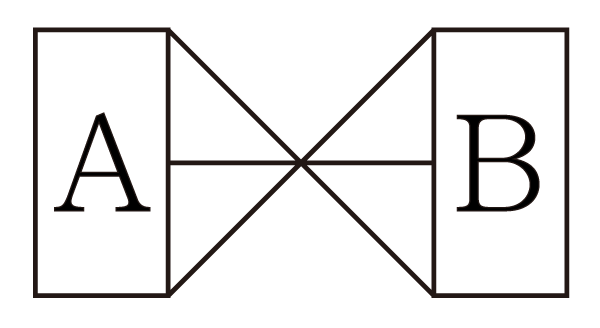}
\end{center}
\caption{\bf 3-pin based NOT gate}\label{Fig06}
\end{figure}

The NOT gate can be operated in a logic-negative operation through one ‘twisting’ as in a 4-pin. To be exact, the position of the middle ground pin is fixed and is a structural transformation that changes the position of the remaining two true and false pins.

\begin{figure}[h]
\begin{center}
\includegraphics[scale=0.1,bb=0 0 600 741]{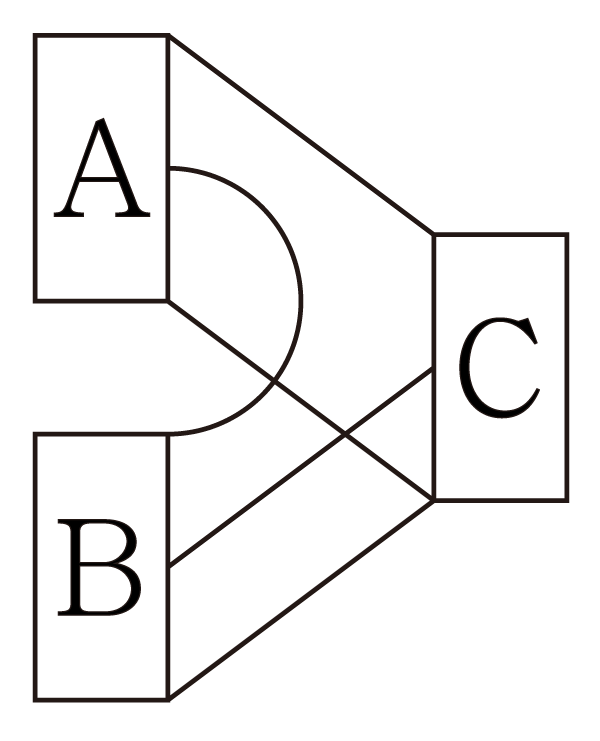}
\end{center}
\caption{\bf 3-pin based AND gate}\label{Fig07}
\end{figure}

\begin{figure}[h]
\begin{center}
\includegraphics[scale=0.1,bb=0 0 600 741]{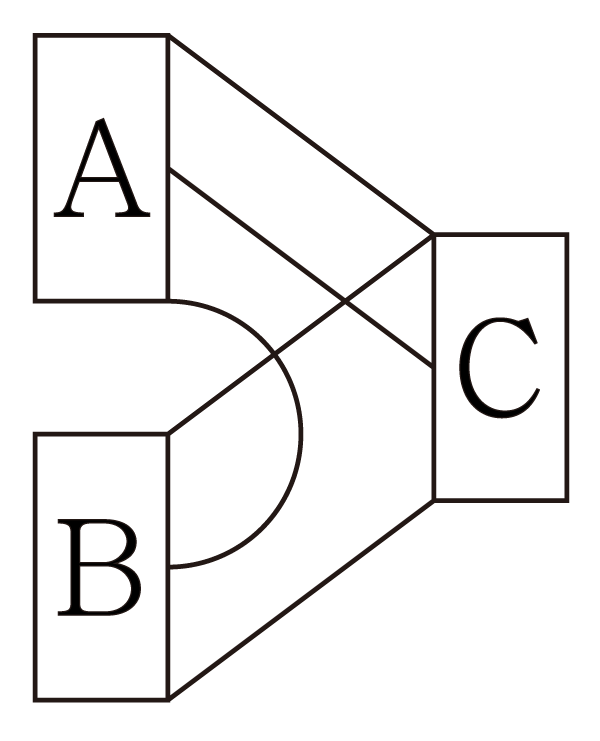}
\end{center}
\caption{\bf 3-pin based OR gate}\label{Fig08}
\end{figure}

\begin{figure}[t]
\begin{center}
\includegraphics[scale=0.2,bb=0 0 400 500]{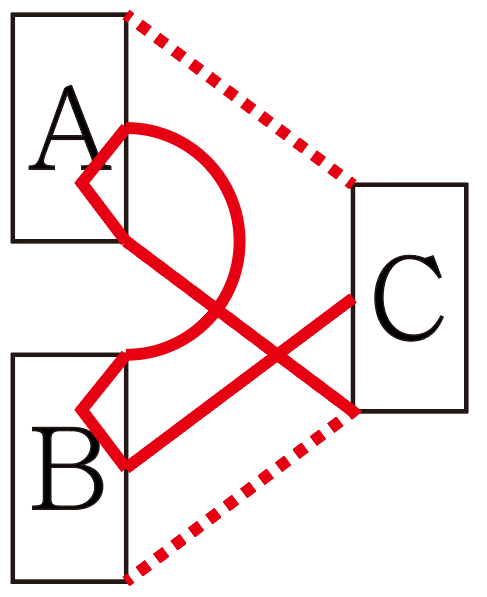}
\end{center}
\caption{\bf 3-pin based AND gate inputs A=0 and B=1}
\label{Fig09}
\end{figure}

\textbf{Fig.~\ref{Fig07}} and \textbf{Fig.~\ref{Fig08}} are AND and/or gate consisting of 3-pin based logics, respectively. \textbf{Fig.~\ref{Fig09}} shows the connection status of the output pin when A=0, B=1 is entered in the AND gate.

As shown in \textbf{Fig.~\ref{Fig09}}, when A=0, B=1, or A is connected, and B is connected, output C is connected only to the following two pins, and this is the correct result for AND operation.

\subsection{Mathematical Meanings: Proof of De Morgan's.}
The basic signal system configuration and 4-pin based, 3-pin based logic gate were described earlier in 2.1 and 2.2. We will then use this paragraph to demonstrate De Morgan's Law of the set theory and explain its mathematical meaning using the signal system of the half-war signal pair.

\begin{theo}
    $\text{About two sets A and B } \\ A^{\mathsf{c}} \cap B^{\mathsf{c}} = (A \cup B)^{\mathsf{c}} \text{(1) or } A^{\mathsf{c}} \cup B^{\mathsf{c}} = (A \cap B)^{\mathsf{c}} \text{(2)} \\ \text{ Satisfies.}$
\end{theo}

First of all, de Morgan's law is of the same mathematical nature as Theorem 1. The general proof of de Morgan's law is shown through the Ben diagram of the truth set. Drawing a Ben diagram shows that, for expressions 1 and 2, the areas of the set represented by the left and right sides match. And truth sets can be expressed not only in a set but also in a logical manner. Describing de Morgan's law as a logical operator is as follows:

\begin{align}
    \text{((NOT(A)) AND (NOT(B)))} &= \text{NOT(A OR B)} \text{ or }\nonumber \\ 
    \text{((NOT(A)) OR (NOT(B)))} &= \text{NOT(A AND B)} \nonumber
\end{align}

Now to prove de Morgan's laws using 3-pin logic systems, we can summarize the structural features of the components of the 3-pin logic system.

\begin{table}[h]
\begin{tabular}{|l|}
\hline
\begin{tabular}[c]{@{}l@{}}\\ 1. Nonlet value: When $\alpha$ line and $\beta$ line cross or mirror, \\ the Nonlet value is inverted. \\ 2. Logicrites: Basically composed of AND, OR gate, \\ AND gate and OR gate are structurally symmetrical. \\ As shown in, this system is commonly changed \\ by 'symmetry'. \\ \\ \end{tabular} \\ \hline
\end{tabular}
\end{table}

To use these properties, assume a sufficiently wide plane mirror. The planar mirror will illuminate any circuit consisting of 3-pin based logic in the specified direction, and this accident experiment will allow us to observe the similarities and differences between the existing circuit and the circuit reflected in the mirror.

\begin{figure}[h]
\begin{center}
\includegraphics[scale=0.25]{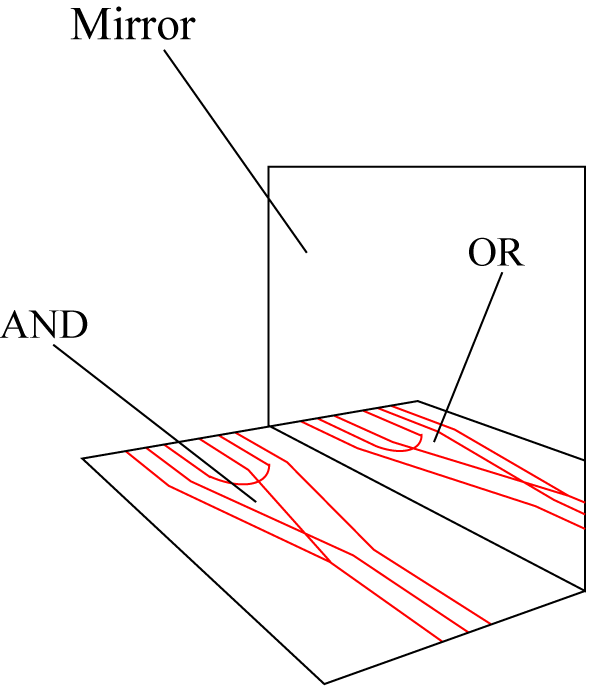}
\end{center}
\caption{\bf De Morgan's Law defined with Structure Based Logic \& Mirror}
\label{mirror_logic}
\end{figure}

First of all, the 3-pin based logic system uses a method of giving input and output using the connection of wires, so the circuitry in the mirror viewed by the experimenter will appear to be working correctly. The shape may be said to be 'circuit in the mirror works correctly' since the connection-non-connections can be clearly defined on the experimenter's basis, even though the existing circuit is shown as a linear representation.

Second, the logical value reflected in the mirror is inverted. Some 3-pin based logic signals have true values (alpha) and inverted values (beta), and the positions of alpha and beta rays change in the circuit in the mirror as viewed by the experimenter. Therefore, the 3-pin based logic signal reflected in the mirror has the opposite signal of the existing signal.

Third, the logical elements in the mirror change each other. Because the AND gate and the OR gate were previously symmetrical, the AND gate in the mirror viewed by the experimenter would be either OR gate, and the OR gate would be seen as AND gate.

According to the above three inference, any 3-pin based logic-based logic circuit in the mirror reverses the input and output values, and changes the AND gate in the circuit to the OR gate and the OR gate to the AND gate. Then any logic circuit will be seen as 

\begin{equation}
    \text{A AND B = C}
\end{equation}

\begin{equation}
    \sim \text{A OR} \sim \text{B} = \sim \text{C}
\end{equation}

in the mirror. At that time, if you substitute (1) for C of (2) , you can get 

\begin{equation}
    \sim \text{A OR} \sim \text{B} = \sim \text{(A AND B)}
\end{equation}

and 

\begin{equation}
    \sim \text{A AND} \sim \text{B} = \sim \text{(A OR B)}
\end{equation}

in the same way. The actual logic circuits work properly, so the circuitry in the mirror also works correctly. In other words, de Morgan's laws can be demonstrated using a 3-pin based logic circuit and a mirror.

\section{Input Control of Logic Circuit}
\subsection{Problems with Logical Replication}
The architectural computers proposed in this paper are capable of logical operations in the logical state of being connected, and their methods are discussed in more detail earlier. There are three inputs, A, B, and C, for some logical operations, e.g. for a full-time generator. Logic values corresponding to A, B, and C are used multiple times, which necessitate the replication of logic. To this end, a split gate of V is defined.

\begin{figure}[h]
\begin{center}
\includegraphics[scale=0.15]{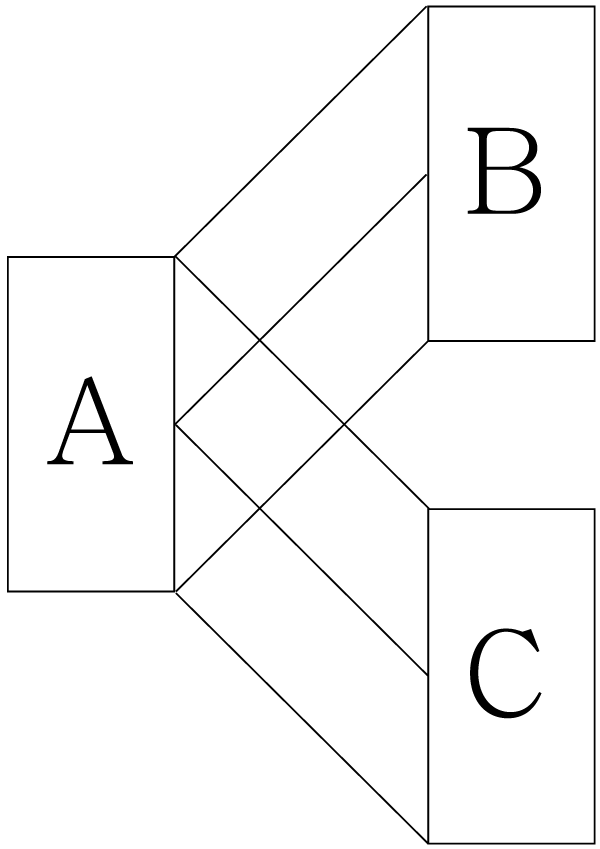}
\end{center}
\caption{\bf split gate}
\label{split_gate}
\end{figure}

However, the splitgate has the risk of reversing current in the process of cloning logic. To solve this problem, we propose a logical replication mechanism using CPU clock.

\subsection{Logical replication using CPU clock}
Counting, one of the basic actions of a computer, is based on a pulse signal with a certain period of time called a CPU clock. CPU clock pulse is a signal that overlaps numerous sine waves through the Fourier transform to form a square wave.

CPU clock pulse, CP (clock pulse) below alternate logic values of 1 and 0 once during cycle T. CPs with these characteristics are defined in the structural computer as repetitive signals, as shown in the graph below.

\begin{figure}[h]
\begin{center}
\includegraphics[scale=0.1]{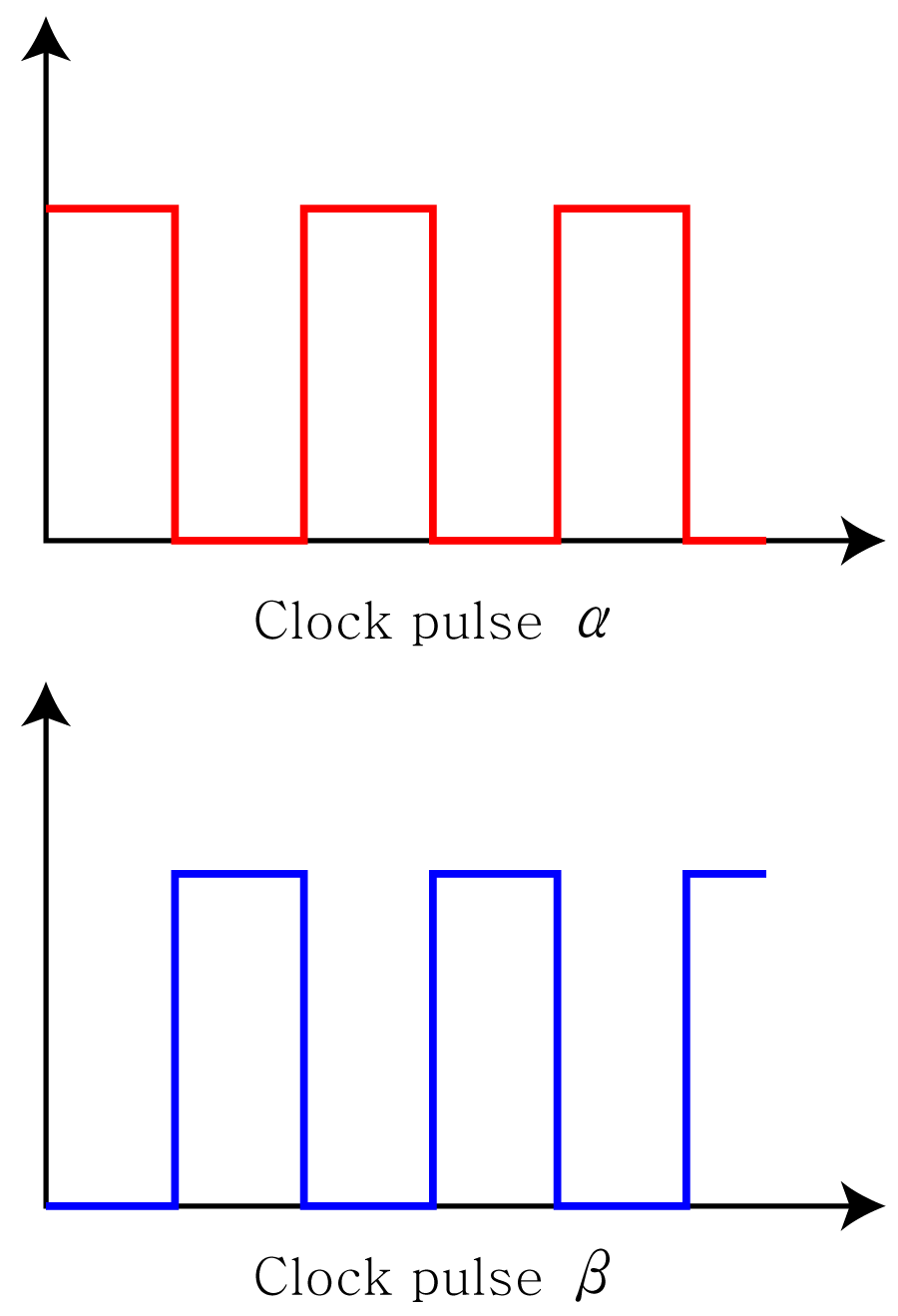}
\end{center}
\caption{\bf Clock Pulse}
\label{clock_pulse}
\end{figure}

\begin{figure}[h]
\begin{center}
\includegraphics[scale=0.15]{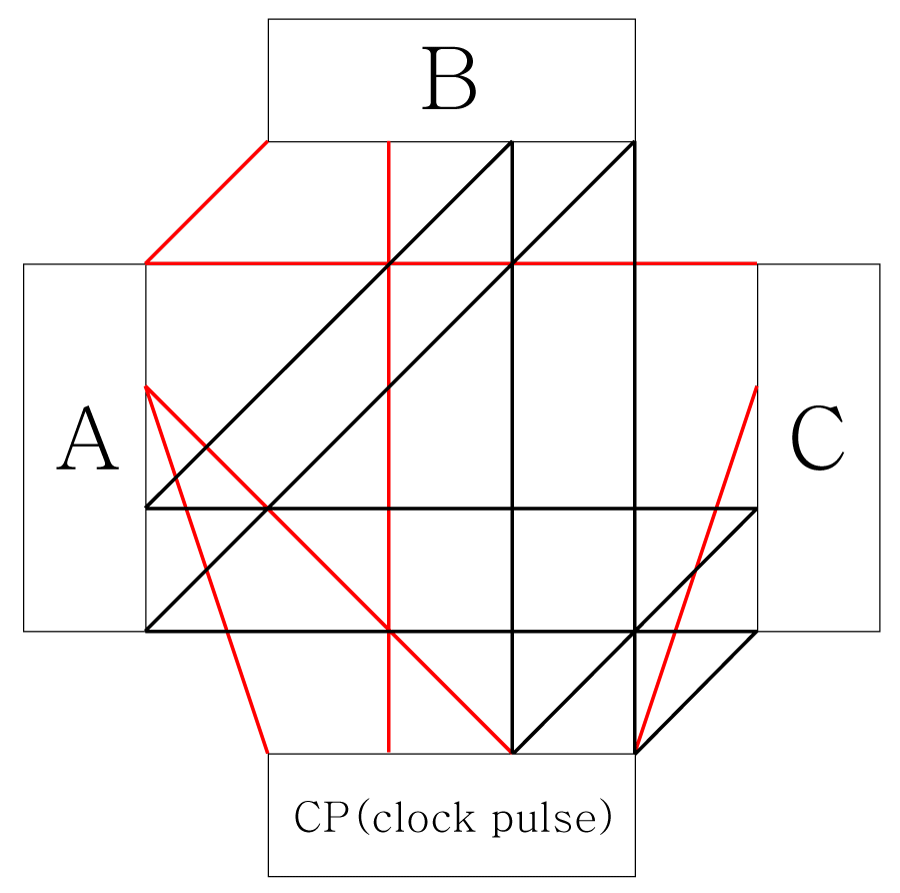}
\end{center}
\caption{\bf Split with Clock Pulse}
\label{split_clockpulse}
\end{figure}

If you use these repetitive signals to redesign the splitgate to replicate one signal independently, same as \textbf{Fig. ~\ref{split_clockpulse}}.

\begin{figure}[h]
\begin{center}
\includegraphics[scale=0.06]{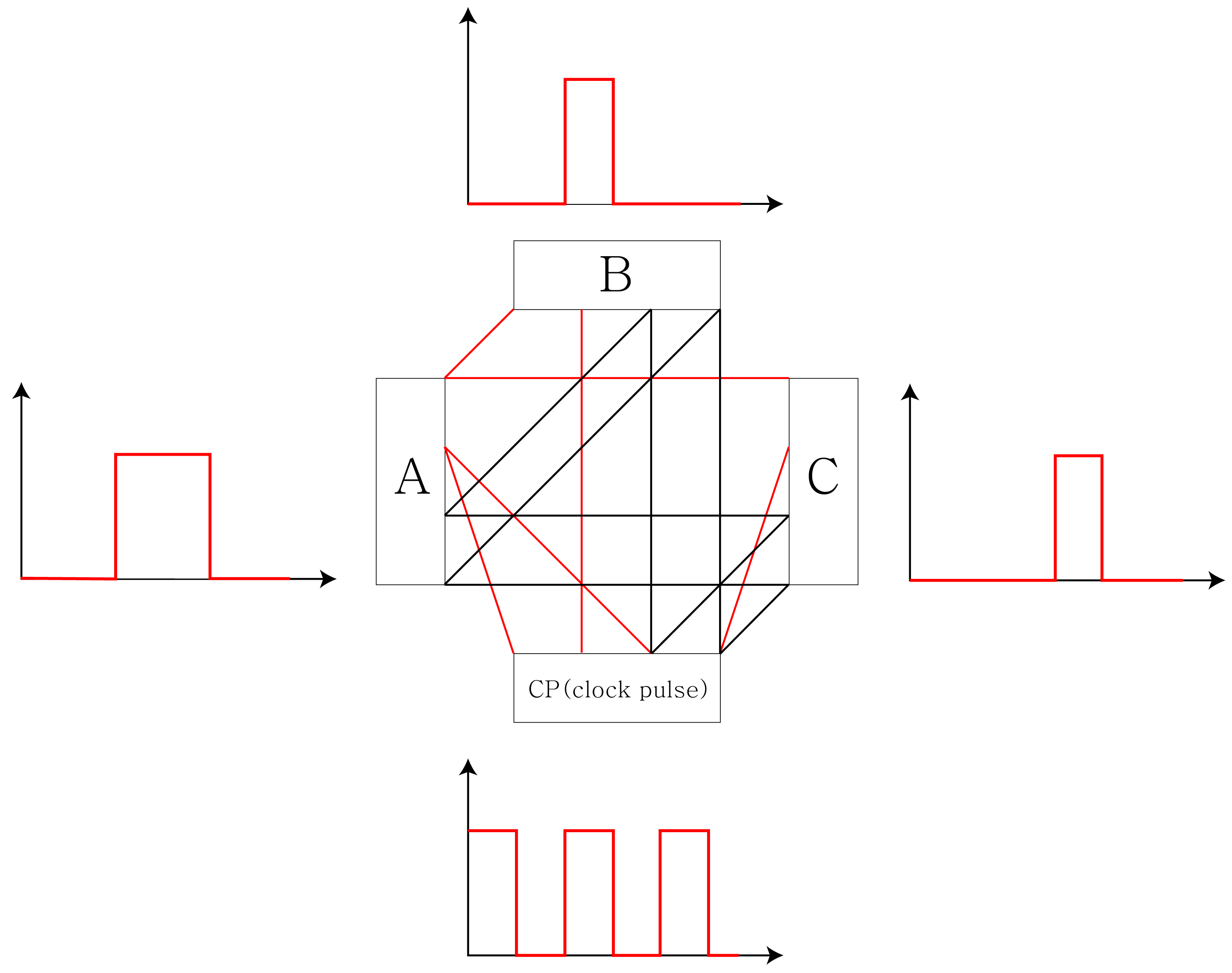}
\end{center}
\caption{\bf Split with Clock Pulse 2}
\label{split_clockpulse2}
\end{figure}

 if you have entered any signal A into the splitgate designed this way. Signal changes, such as \textbf{Fig~\ref{split_clockpulse2}}.

\begin{figure}[h]
\begin{center}
\includegraphics[scale=0.1]{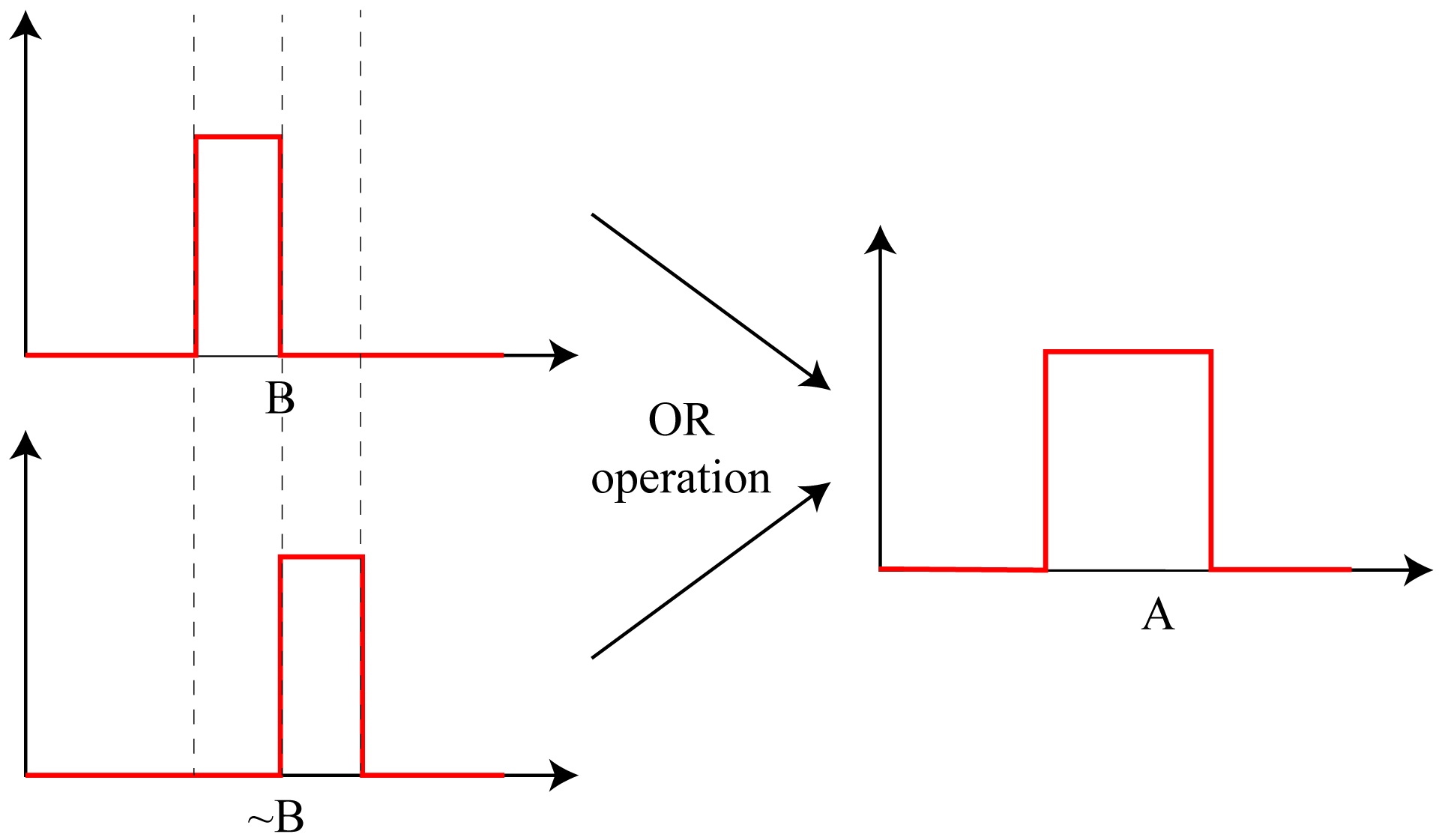}
\end{center}
\caption{\bf Split Final Process}
\label{split_final_process}
\end{figure}

To replicate a signal A, preserve the alpha signal, beta signal, and reverse signal pair in the CP and calculate the AND respectively. Computed partial signals are same as B and C like \textbf{Fig.  ~\ref{split_clockpulse2}}. The OR operation of the primary and reverse signals on these B and C signals will allow you to recover the partially maintained signal and the schematic diagram of the process is same as \textbf{Fig. ~\ref{split_final_process}}.

This method allows logical replication using CP to solve the problem of back-current generation of conventional parallel connections.

\section{Simulation with Depth-First Search}
\begin{figure}[h]
\begin{center}
\includegraphics[scale=0.08,bb=0 0 2048 1536]{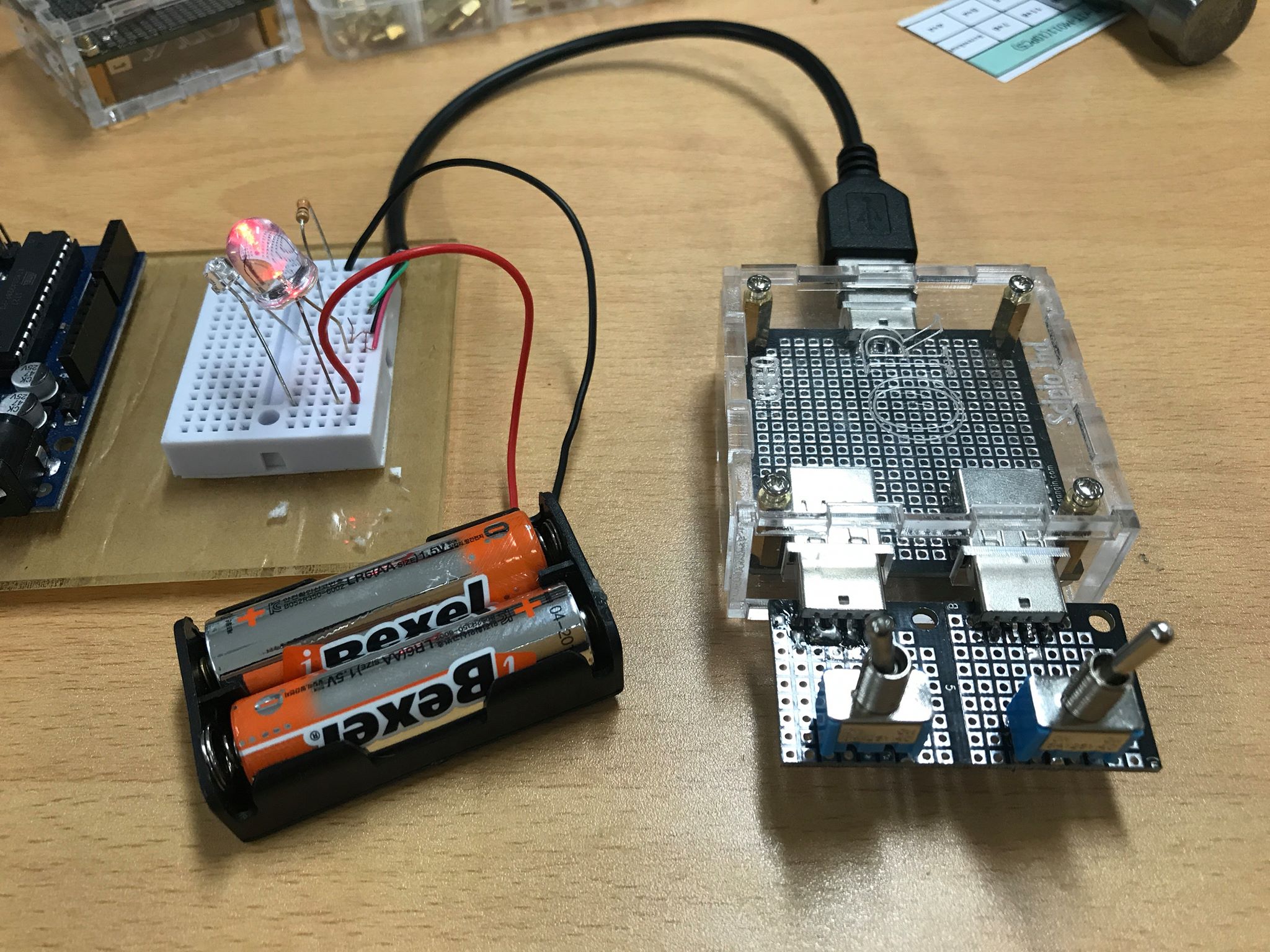}
\end{center}
\caption{\bf USB Experiment OR Gate}\label{usbor}
\end{figure}

To simulate the aforementioned structural computer theory, a device in the form of a USB connection \textbf{Fig.~\ref{usbor}}, created with \textbf{Fig. ~\ref{usbcircuit}}. However, as the circuit grows in size, a number of USB-connected simulation devices are required, resulting in cost problems.

\begin{figure*}[t]
\begin{center}
\includegraphics[scale=0.33]{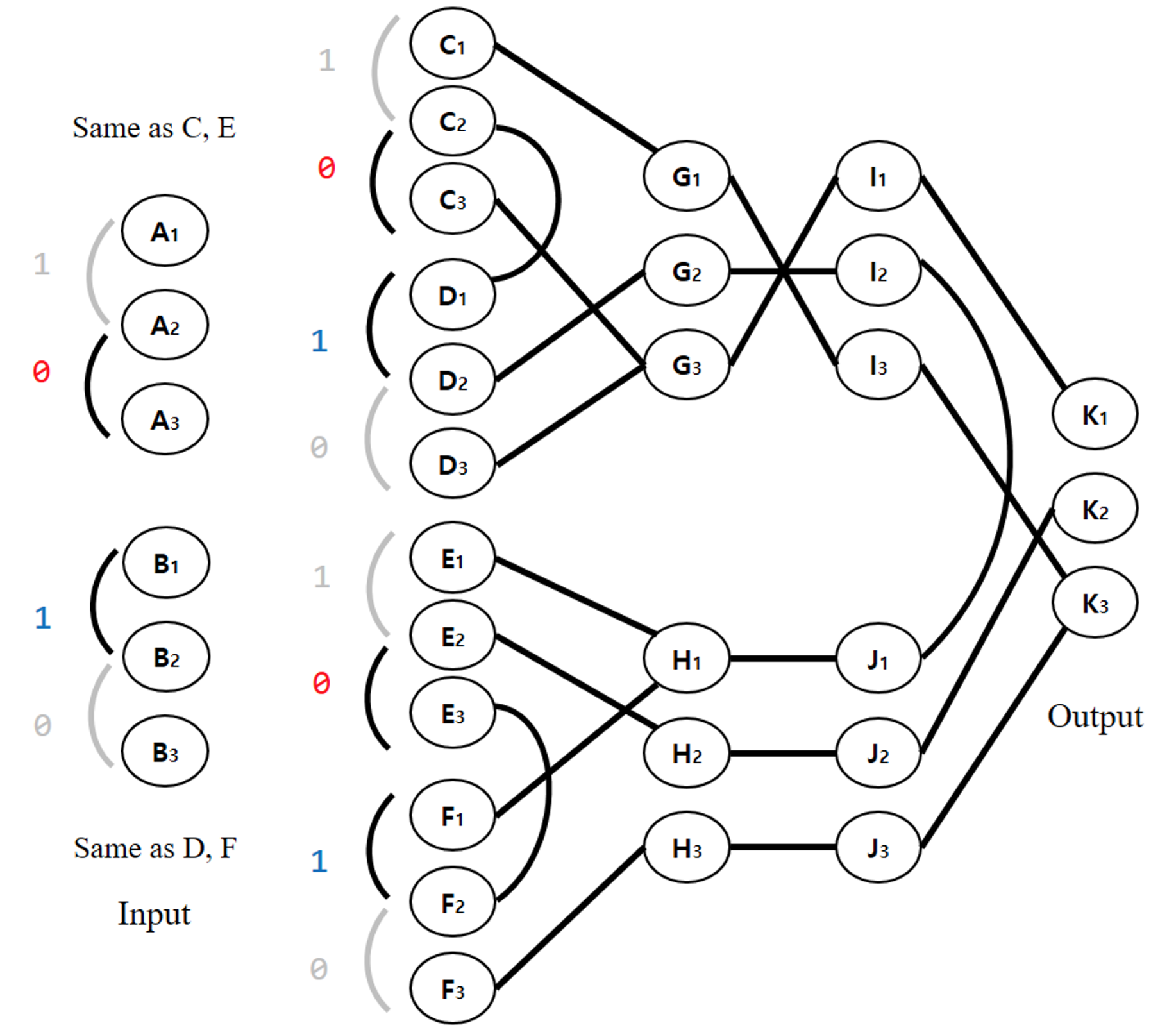}
\end{center}
\caption{\bf Example of XOR((A NAND B) AND (A OR B)) Gate into Vertex and Edge Graph}\label{dfs_nandor}
\end{figure*}

\begin{figure}[h]
\begin{center}
\includegraphics[scale=0.12]{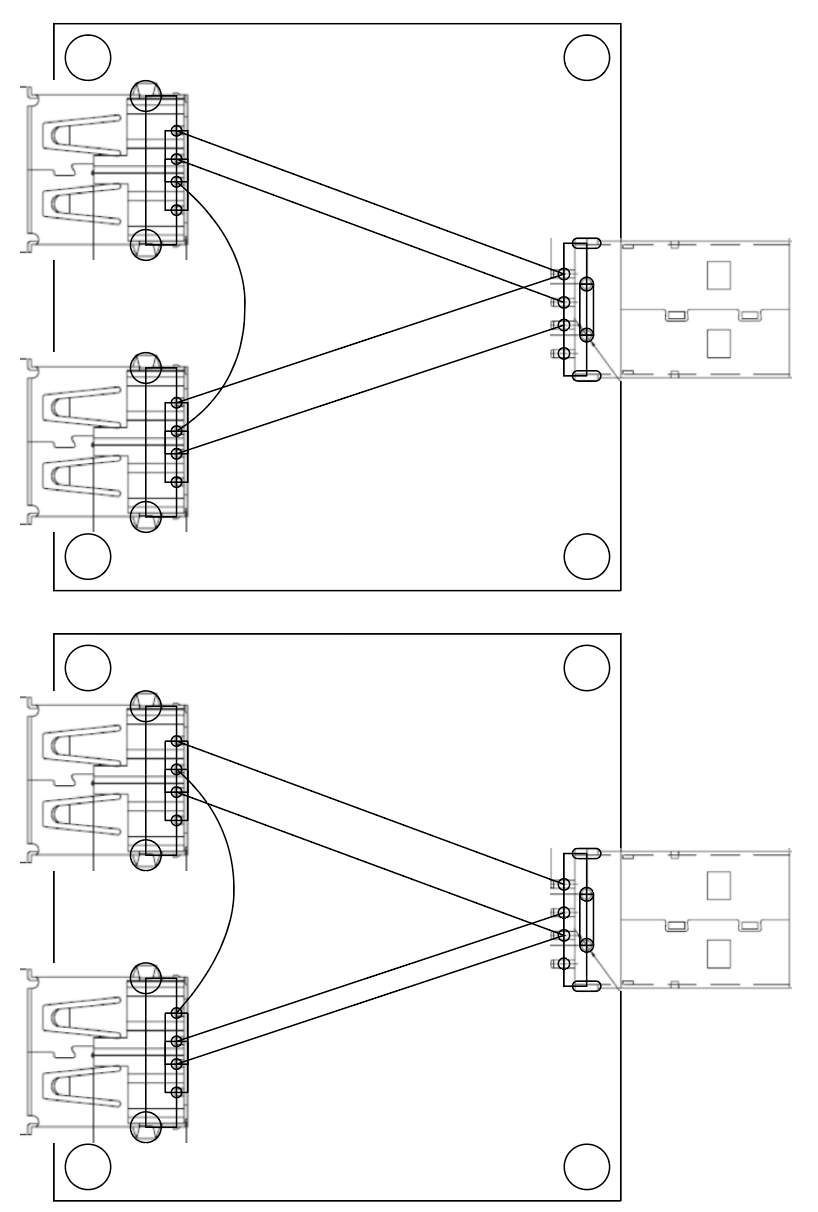}
\end{center}
\caption{\bf USB Experiment Gate Circuit}\label{usbcircuit}
\end{figure}

We decided to verify that the structural computer theory presented so far is actually working without the cost of circuit building, to simulate the connection of complex Circuits rather than just Gate Circuit, and to set up Metric for experiments that can test structural computers for logical errors and error.

Depth-First Search (DFS) is a method of starting with the Root Node on that Graph and fully exploring that branch before moving on to the next Branch, when any Vertx-Edge based Graph is entered in one of the nonlinear data structures. Having the form of a circular algorithm calling itself, it is implemented as a recursive function, and other forms of tree rotation, including potential rotation, are all considered types of DFS.

For graph navigation, Stack data structures can be used because it must be examined which nodes were visited. DFS is defined as two Function as \textbf{Algorithm 1}.

\begin{algorithm}[h]
\caption{Depth-first Search}\label{euclid}
\begin{algorithmic}[1]
\Function{DFS}{G}
\ForAll{vertex u}
	\State color[u]=red
	\State parent[u]=-1
	\State time=0
\EndFor
\ForAll{vertex u}
\If{color[u]==red}
\State DFS-Visit(u)
\EndIf
\EndFor
\EndFunction
\end{algorithmic}

and

\begin{algorithmic}[1]
\Function{DFS-Visit}{u}
\State color[u]=gray
\State time=time+1
\State d[u]=time
\ForAll{v $\in$ adj[u]}
\If{color[v]==red}
\State parent[v]=u
\State DFS-Visit(v)
\EndIf
\EndFor
\State color[u]=black
\State time=time+1
\State f[u]=time
\EndFunction
\end{algorithmic}
\end{algorithm}

\begin{table}[h]
\centering
\begin{tabular}{|c|c|}
\hline
Start Vertex & End Vertex \\ \hline
C1,E1           & K1         \\ \hline
C1,E1           & K2         \\ \hline
C1,E1           & K3         \\ \hline
C2,E2           & K1         \\ \hline
C2,E2           & K2         \\ \hline
C2,E2           & K3         \\ \hline
C3,E3           & K1         \\ \hline
C3,E3           & K2         \\ \hline
C3,E3           & K3         \\ \hline
D1,F1           & K1         \\ \hline
D1,F1           & K2         \\ \hline
D1,F1           & K3         \\ \hline
D2,F2           & K1         \\ \hline
D2,F2           & K2         \\ \hline
D2,F2           & K3         \\ \hline
D3,F3           & K1         \\ \hline
D3,F3           & K2         \\ \hline
D3,F3           & K3         \\ \hline
\end{tabular}
\caption{Start Vertex and End Vertex}
\label{dfs_startend}
\end{table}

\begin{table}[h]
\centering
\begin{tabular}{|c|c|c|}
\hline
Start Vertex & End Vertex & Result \\ \hline
C1,E1           & K1         & O         \\ \hline
C1,E1           & K2         & X         \\ \hline
C1,E1           & K3         & O         \\ \hline
C2,E2           & K1         & O         \\ \hline
C2,E2           & K2         & X         \\ \hline
C2,E2           & K3         & X         \\ \hline
C3,E3           & K1         & O         \\ \hline
C3,E3           & K2         & X         \\ \hline
C3,E3           & K3         & X         \\ \hline
D1,F1           & K1         & O         \\ \hline
D1,F1           & K2         & X         \\ \hline
D1,F1           & K3         & X         \\ \hline
D2,F2           & K1         & O         \\ \hline
D2,F2           & K2         & X         \\ \hline
D2,F2           & K3         & X         \\ \hline
D3,F3           & K1         & O         \\ \hline
D3,F3           & K2         & X         \\ \hline
D3,F3           & K3         & O         \\ \hline
\end{tabular}
\caption{Result of Start Vertex and End Vertex test case}
\label{dfs_startend_result}
\end{table}

\subsection{Development of S/W using C Programming}
Using the above DFS concept, the C programming code was written and used in the experiment, and the code produced is as follows.

\begin{lstlisting}[style=CStyle]
#include <stdio.h>
#include <stack>
using namespace std;
stack<int> st;
int n, m, endd, map[100][100], visited[100];
char arr[11] = {'A', 'B', 'C', 'D', 'E', 'F', 'G', 'H', 'I', 'J', 'K'};

void visit(int a) {
    visited[a]=1;
    st.push(a);
}

void dfs(char k) {
    int i;
    visit(k);
    while(!st.empty()) {
        if(st.top()==endd) return;
        for(i = 1; i <= n; i++) {
            if(map[st.top()][i] && !visited[i]) {
                visit(i);
                break;
            }
        }
        if(i == n+1) st.pop();
    }
}

void print() {
    if(st.empty()) return;
    int ans = st.top();
    st.pop();
    print();
    printf("%c%d ", arr[(ans-1)/3], (ans%3==0 ? 3 : ans%3));
}

int main() {
    scanf("%d %d", &n, &m);
    for(int i = 0; i < m; i++) {
        int v1, v2;
        scanf("%d %d", &v1, &v2);
        map[v1][v2] = 1;
    }
    int start;
    scanf("%d %d", &start, &endd);
    dfs(start);
    print();
    return 0;
}
\end{lstlisting}

In this programming code, a vertex-Edge graph is entered and expressed as a directional adjacent matrix. An directionless adjacent matrix is implemented as a graph that allows two vertices to be freely navigated using the trunk, while a directional adjacent matrix is implemented as a graph that can be explored only with a specified path between the two vertex points.

Insert the Vertex-Edge graph \textbf{Fig.~\ref{dfs_nandor}} into our code and results as shown in \textbf{Table.~\ref{dfs_inout_result}}.

\begin{table*}[h]
\centering
\begin{tabular}{|c|c|c|c|}
\hline
Start \& End & A & B & Result                                                \\ \hline
K2 $\rightarrow$ K1      & 0 & 0 & X                                                     \\ \hline
K2 $\rightarrow$ K1      & 1 & 0 & K2 J2 H2 E2 E1 H1 J1 I2 G2 D2 D3 G3 I1 K1                         \\ \hline
K2 $\rightarrow$ K1      & 0 & 1 & K2 J2 H2 E2 E3 F2 F1 H1 J1 I2 G2 D2 D1 C2 C3 G3 I1 K1 \\ \hline
K2 $\rightarrow$ K1      & 1 & 1 & X                         \\ \hline
K2 $\rightarrow$ K3      & 0 & 0 & K2 J2 H2 E2 E3 F2 F3 H3 J3 K3                         \\ \hline
K2 $\rightarrow$ K3      & 1 & 0 & X                                                     \\ \hline
K2 $\rightarrow$ K3      & 0 & 1 & X                         \\ \hline
K2 $\rightarrow$ K3      & 1 & 1 & K2 J2 H2 E2 E1 H1 J1 I2 G2 D2 D1 C2 C1 G1 I3 K3             \\ \hline
\end{tabular}
\caption{Result of Input and Output test case}
\label{dfs_inout_result}
\end{table*}

\subsection{DFS(Depth-First Search) Verification Simulation}
Graph described in \textbf{Fig. ~\ref{dfs_nandor}} is an implementation of an XOR gate combining NAND and OR, expressed in 33 vertices and 46 mains. Graphs are expressed in red and blue numbers in cases where there is no direction of the main line (the main line that can be passed in both directions) and the direction of the main line (the main line that can only be moved outward from the middle of the set of vertex).

First, go through the Test Evaluation process to confirm that this circuit is correctly designed. Because of the 3-pin input, A2 and A1, B2 and B1 can be expressed as 1 respectively, and vice versa, A2 to A3 and B2 to B3. This input is directional, so both A and B Inputs start at A2 and connect to A1 or A3. On the contrary, it is impossible to navigate. The current A vertex (A1, A2, A3) input is expressed as 0 because A2 and A3 are connected, and the B vertex (B1, B2, B3) input is expressed as 1 because B2 and B1 are connected.

Inputs from A vertex (A1, A2 and A3) sets and from B vertex (B1, B2 and B3) sets are processed the same as input from D and F vertex sets, \textbf{Table.~\ref{dfs_startend}} and are also \textbf{Table.~\ref{dfs_startend_result}}. To verify that this input is acceptable to the circuit, experiment with the input phase A and B vertex (6), the output phase K vertex (3), and a total of 18 test cases. This is the same as \textbf{Table.~\ref{dfs_startend}}.

Subsequently, DFS (Depth First Search) verifies that the output is possible for the actual Pin connection state. As described above, the output is determined by the 3-pin input, so we will enter 1 with the A2 and A1 connections, the B2 and B1 connections (the reverse is treated as 0), and the corresponding output will be recognized through DFS navigation. In this course, we experiment with a total of eight test cases, including the number of input branches (four) of XOR and the direction of mobility of the output pin (K1 in K2 and K3 in K2).

\subsection{DFS (Depth First) Verification Simulation Results}
We will look at the inputs through 18 test cases to see if the circuit is acceptable. Results \textbf{Table.~\ref{dfs_startend_result}} As shown in , searchability is expressed in 'O' and non-searchable 'X'. The experiment was based on the same vertex-Edge Graph as \textbf{Fig. ~\ref{dfs_nandor}} so it can be seen that this circuit is logically operational.

Next, it verifies with DFS that the output is possible for the actual pin connection state. As mentioned above, the search is carried out and the results are expressed by the unique number of each vertex. The result is as shown in \textbf{Table.~\ref{dfs_inout_result}}. The result of moving from the K2 peak to the K1 peak is the same as that of the XNOR, and the result of moving from the K2 peak to the K3 peak is the same as that of the XOR, it is possible to confirm that this study is feasible.

\section{Address the Problems of Semiconductor}
\subsection{Address Heat and Heat Loss Problems}
The biggest problem with most semiconductor-based electronic circuits right now is that they have a high fever. This is why the heat shield attached to the integrated circuit and the cooling fan of the computer are needed.

The causes of heat losses in semiconductors can be divided into three main reasons. First, semiconductors have self-resistance because they are a single electrical circuit in themselves. However, any electronic circuit has basic electrical resistance and its size is very small, so it is hard to judge that it is a problem only with semiconductors.

Second, the loss of power produced by the voltage drop in the current channel junction layer of junction elements such as transistors. For semiconductors, the moment the electromotive force reaches a certain voltage, it becomes possible to have all current values. This is different from resistance and is the characteristic of diodes with a constant voltage drop regardless of how much current is flowing. Because of this nature, diodes have a power consumption that is directly proportional to the current, not the square of the current, as opposed to the resistance.

\begin{figure}[t]
\begin{center}
\includegraphics[scale=0.8]{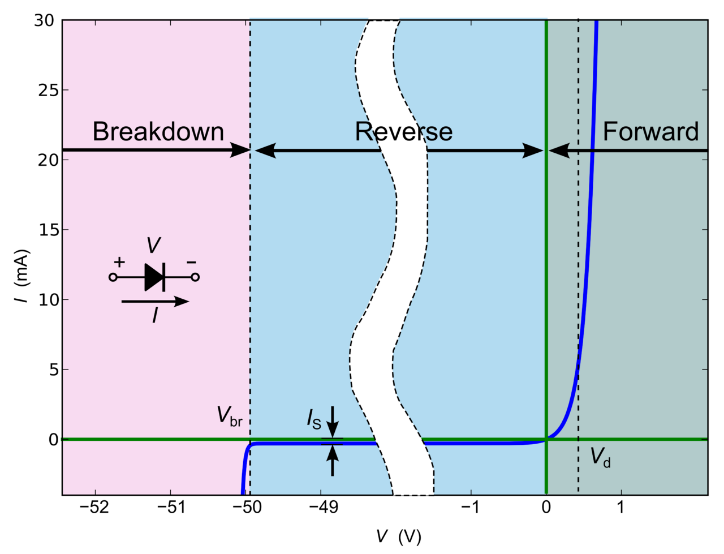}
\end{center}
\caption{\bf Graph of current allowed in diodes \cite{rc}}\label{rc_figure}
\end{figure}

\begin{figure}[t]
\begin{center}
\includegraphics[scale=0.4]{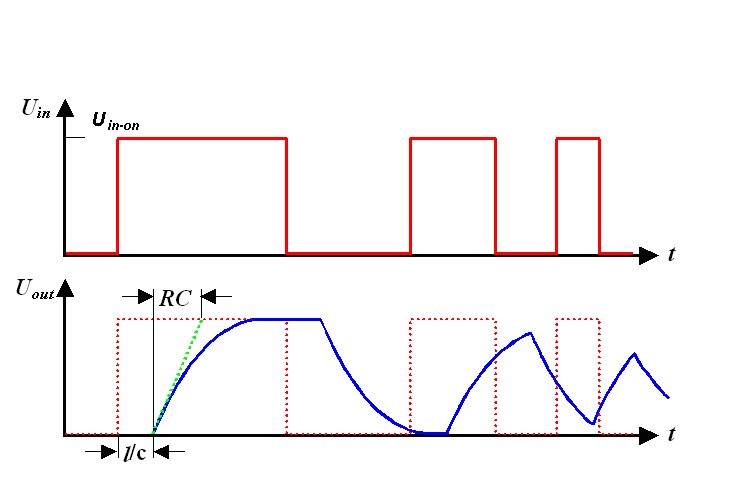}
\end{center}
\caption{\bf Figure of RC-Delay \cite{rc2}}\label{rc2_figure}
\end{figure}


Therefore, it can be seen that the heat output of the integrated circuit is proportional to the strength of the current, the voltage drop of the semiconductor, and the number of semiconductor devices. However, current transistor-based circuits use transistors for logic operations as well as switching operations. The number of transistors required for switching operations can be considered linear, as it is proportional to the number of input signals, but the number of transistors used for logical aggregates can be seen to be relatively large.

By contrast, structural computers do not use transistors in their logic, requiring only a linear number of devices to switch input signals.

For example, considering the AND gate, for semiconductor transistors, two transistors are required to give input signals, two transistors are required for the computation process, and a total of four elements, but only two devices for input are required for structural computers.

More generally, let's define the number of elements required in a semiconductor integrated circuit that goes through $k$ simple logical operation, $i$ NOT operation using $n$ input signals as a function of $f(n,k,i)$. Simple logic operation means AND and OR. Simple logic operations require two transistors and one transistor for NOT operations, which can be described as follows:

\begin{equation}
    f(n,k,i)=n+2k+i
\end{equation}

If the number of devices required to implement the same situation into a structural computer is $g(n,k,i)$
\begin{equation}
    g(n,k,i)=n
\end{equation}

This is because it is $n<<k,i$ for integrated circuits of a certain size or larger, so for $f$ and $g$

\begin{equation}
    for \text{ } n << k,i \Rightarrow g(n,k,i) << f(n,k,i)
\end{equation}

it can be said that In other words, semiconductor computers use a lot more junction devices than structural computers and emit proportional heat. Therefore, the use of structural computers will reduce heat generation problems, which are problems with conventional semiconductors, to prevent aging and damage.

\subsection{Decrease in propagation delay and increase in computational speed}
Every RC circuit has a physical phenomenon called RC delay. RC delay is a phenomenon in which pulses signals, such as right-angled waves, are externally deformed by a capacitor in a circuit, and due to the nature of semiconductors that require more than a certain size of electromotive force to flow, RC delay is a major problem that reduces the accuracy of logic operations.

The effect of RC delays cannot be eliminated because the circuit itself can act as a accumulator even if it is not used separately in the circuit. This phenomenon in an integrated circuit is called Projection Delay, and the delay is the time constant of the RC circuit. \\
\begin{equation}
    \tau =RC
\end{equation}

be known as proportional to For semiconductors, the accumulated resistance across the circuit increases due to ground resistance and its own resistance. For the electrical capacity of a circuit, it is known as a physical quantity that does not change much unless the size of the circuit varies greatly. Thus, semiconductor-based computers will have a large time constant ($\tau$) at the point of circuit theory and a relatively long Projection Delay phenomenon will occur.

On the other hand, in the case of structural-based computers, the self-resistance and ground resistance of semiconductors do not exist by not using semiconductor devices, and only the resistance of the wires themselves remains. Because the resistance of conductors is very small, the resistance of the circuit required to do the same logical operation will be lower than the aforementioned semiconductor-based computers, and will have a lower time constant value. Thus, a structure-based computer will experience a shorter Projection Delay phenomenon than a semiconductor-based computer. This will allow the architecture-based computer to have faster computational speeds and, in ordinary cases, faster CPU clocks than conventional semiconductor-based computers.

\section{Applications}
\subsection{Internet Communication Physical Control: Serial Communication Experiment using Arduino}
Internet communication refers to the process of transferring data from any computer to another computer. Internet communication uses TCP and UDP protocols, which transmit data after confirming that a connection is made between the sending and receiving sides, and UDP protocols unilaterally transmit data without going through the verification process of the connection between the receiving end and the sending side.

If the data is sent using a TCP protocol or UDP protocol, use the TCP header, IP header, and routing table. In this experiment, it is shown through Arduino's serial communication that the principles of the structural computer in wired Internet communication can be used to send and receive information to and from multiple PCs without various headers and routing tables.

\begin{figure}[h]
\begin{center}
\includegraphics[scale=0.05]{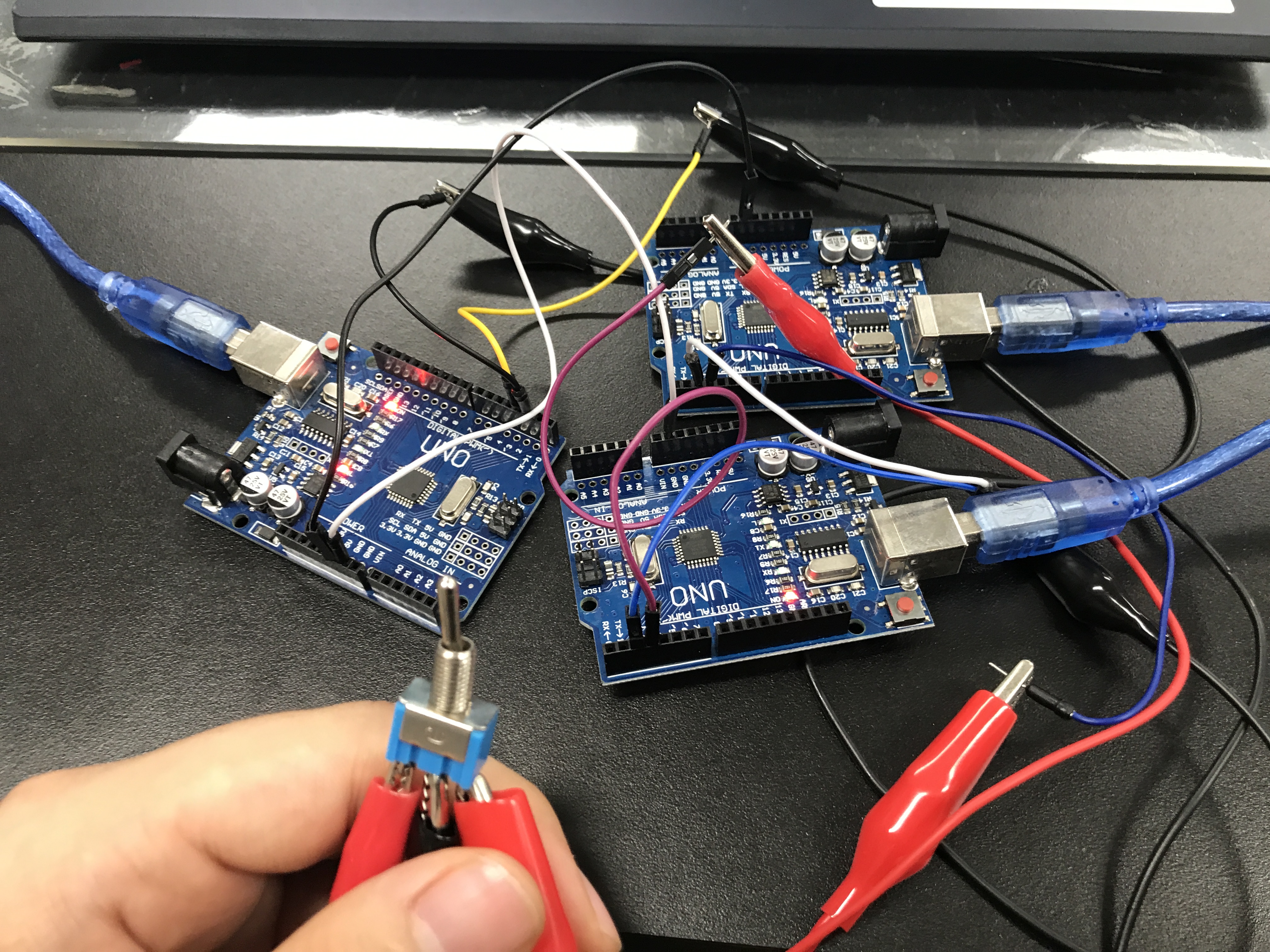}
\end{center}
\caption{\bf Experiment of Ethernet Communication using 3 arduino}
\label{ethernet_experiment}
\end{figure}

The experimental environment is as follows and consists of the \textbf{Fig.~\ref{ethernet_experiment}}.

\begin{table}[h]
\begin{tabular}{|l|}
\hline
\begin{tabular}[c]{@{}l@{}} \\
1. Experiment is carried out using three arduinos.\\
2. Conduct an experiment using the principles of \\ the structural computer and determine the \\ direction of data transmission through the switch.\\
3. Using serial communication, check if sending and \\ receiving is possible without a error from both sides.\\ \\ \end{tabular} \\ \hline
\end{tabular}
\end{table}

The source code produced for the experiment is as follows. For each arduino, the following source code was compiled and uploaded and an experiment was carried out.

\begin{lstlisting}[style=CStyle]
#include <SoftwareSerial.h>
SoftwareSerial mySerial(2, 3);

void setup() {
  Serial.begin(9600);
  mySerial.begin(9600);
}

void loop() {
  if(mySerial.available()) {
    Serial.write(mySerial.read());
  }
  if(Serial.available()) {
    mySerial.write(Serial.read());
  }
}
\end{lstlisting}

Through the above source code, the transmission direction from a particular arduino to a particular arduino was switched, and the message was confirmed to be sent by hand.

\subsection{Easy production of IC chips using 3D printing}
With the recent adoption of 3D printing technology, users have begun 3D printing objects of various materials and shapes. Color 3D printing, which uses different colors of filaments alternately rather than just one kind of plastic filament, and metal 3D printing technology, which creates structures by melting metal powder momentarily.

3D printing is basically derived from a technology called Computer Numerical Control (CCC) that controls motors quantitatively. The three-axis orthogonal CNC, the most common type of CNC, controls the X, Y and Z axis coordinates by moving the three step motors, controlling the machine coordinates of the CNC's nozzle or processor.

A step motor is a type of DC motor that uses electromagnetic induction and is driven in such a way that the direction of the axis is determined by the magnetic force of the coil that encloses the motor shaft inside, unlike a conventional direct current motor with continuous movement. The step motor operates in relation to the minimum operating unit called step and its precision depends on the number of coils in the step motor.

Because signals for control are complex compared to DC motors, they are usually used with step motor drivers, but are more precise than DC motors in terms of specifying the position of the motor axis with magnetic force, which is why they are used on CNCs.

Most CNCs, including the 3D printer, are commanded in a language called g-code. G-code is a language designed to divide CNC's movements into basic units so that users can fully control CNC's movements with minimal commands. For example, in terms of ease of production and processing, a structure-based computer is very suitable for production in 3D printing. This is because unlike current computers, the architecture-based computer is designed to enable logical operation only with the placement of wires without the use of special devices such as semiconductor transistors. The g-code divides the complex shapes into fine lines, compressing commands, and the architecture-based computer will be able to produce ICs (integrated circuits) with only a simple operation because a single logical gate consists of no more than 10 wires. The g-code for producing the AND gate of a structure-based computer is as follows.

\begin{figure}[h]
\begin{center}
\includegraphics[scale=0.8]{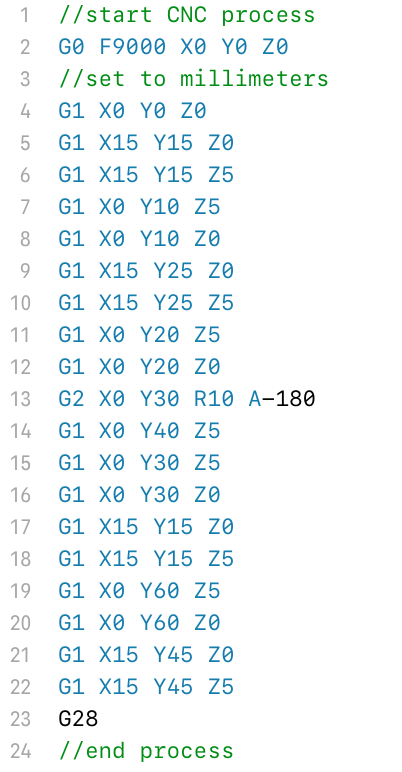}
\end{center}
\caption{\bf Example of G-Code}
\label{gcode}
\end{figure}

\begin{figure}[h]
\begin{center}
\includegraphics[scale=0.15]{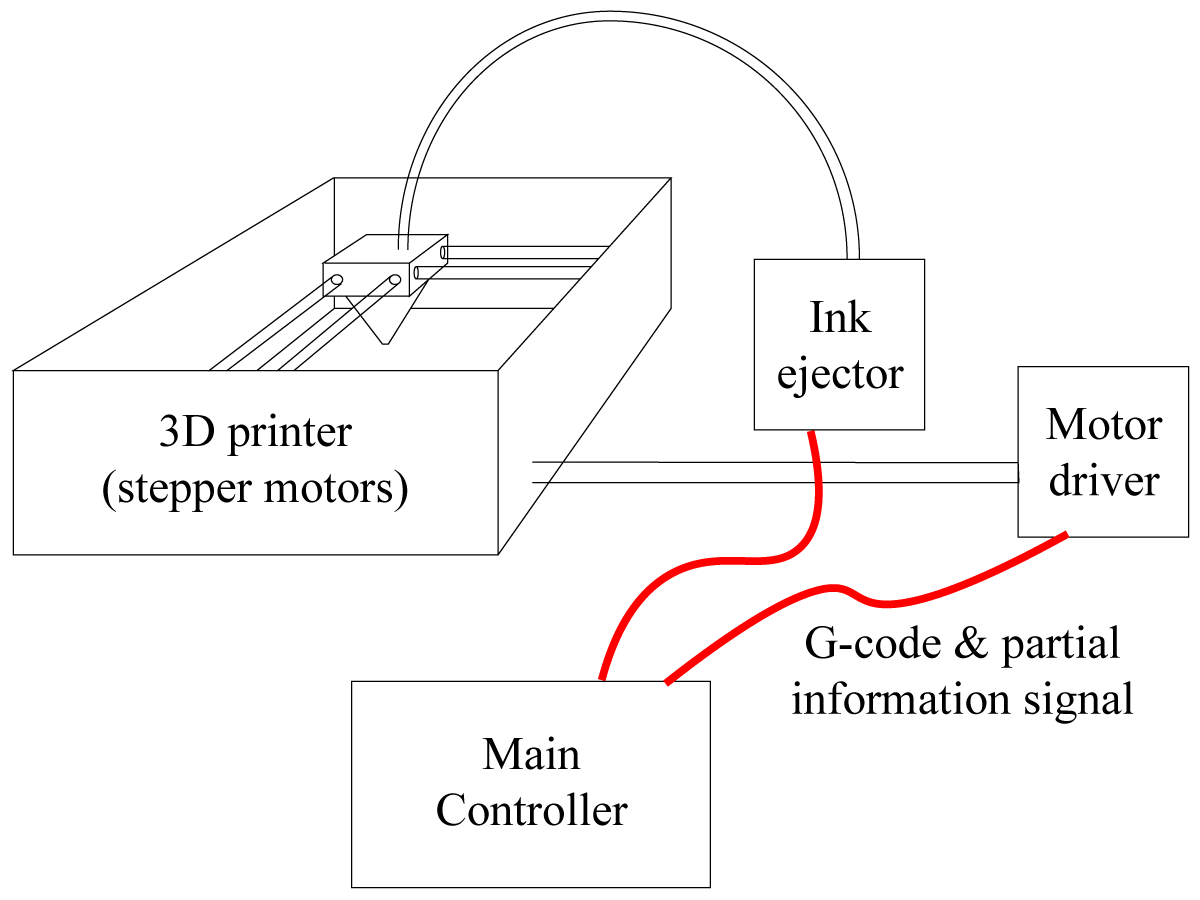}
\end{center}
\caption{\bf IC chip producing using 3D printers}
\label{3dprinter_logic}
\end{figure}

If the IC circuit is produced using the microprocess of the 3D printer, conductive ink (ink) containing electrically conductive materials, such as silver, may be sprayed on the non-conductor board. Typical conductive ink points include conductive pens sold on the market. The embodiment of the Mechanism is shown in the \textbf{Fig.~\ref{3dprinter_logic}}.

Using this method, unlike current semiconductors, integrated circuits can be produced without having to go through complicated processing. Current semiconductor-based computers use lasers and etchings on silicon wafers to count circuits, limiting substrate materials to multi-stage processes, high costs, and ‘wafers’.

On the other hand, if a structure-based computer is 3D printed, it can produce integrated circuits cheaply in a single-stage process, and whatever the material of the circuit board is, is possible.

\section{Conclusion}
This paper presents the NOT gate implementation of structural computers and the Reverse-Logic pair and double pair-based logic operation techniques of digital signals that can solve the problem of heating and aging of existing semiconductor computers.

Unlike conventional silicon-based semiconductors that use resistors and transistors together, structural computers that do not use resistance and are made up of twisted wires are expected to significantly reduce heat generation than current computers, thus improving the computer's computational speed (clocks).

To implement the above idea, this paper has built a prototype of a USB-type AND and OR gate structure computer, and proposes a logic replication mechanism using CPU Clock to prevent the risk of current being reversed by the replication of logic values in the structural computer. Related to the mechanism of the Fast Fourier Transformation algorithm, stable logic replication was implemented by utilizing a system to recover partial signals again.

Furthermore, we propose Simulation Metric (DFS) based on deep-first search (DFS) that enables easy implementation and testing of complex structural computer Circuits. This confirmed the feasibility of this study in an experiment based on an XOR gate produced by combining NAND, AND and OR gates.

In addition, this research has been verified in the paper through experiments and theories that the value of utilizing the physical control of the wired Internet and the ease of production of IC chips using 3D printing.

And it is expected that this research can be applied to the development of artificial intelligence technologies such as deep learning in the future. In other words, it is expected that the idea of structural computers will be applied to semiconductors that generate a lot of heat, such as GPU servers that require processing of large amounts of data, to drastically reduce heat generation, reduce electricity use, and improve performance more than before.


\end{document}